\documentclass{article}
\usepackage{spconf,amsmath,graphicx}
\usepackage[dvipsnames]{xcolor}
\usepackage{enumitem}
\usepackage{booktabs}
\usepackage{threeparttable}
\usepackage[hang,flushmargin]{footmisc}
\usepackage{hyperref}

\title{A Sidecar Separator Can Convert a Single-Talker Speech Recognition System to a Multi-Talker One\vspace{-0.1cm}}

\name{Lingwei Meng, Jiawen Kang, Mingyu Cui, Yuejiao Wang, Xixin Wu, Helen Meng\vspace{-0.1cm}}
\address{The Chinese University of Hong Kong, Hong Kong SAR, China\\\texttt{\ninept{\{lmeng, jwkang, mycui, wangy, wuxx, hmmeng\}@se.cuhk.edu.hk}}\vspace{-0.1cm}}
\begin{document}
\ninept

\maketitle

\begin{abstract}
\vspace{0.2cm}

Although automatic speech recognition (ASR) can perform well in common non-overlapping environments, sustaining performance in multi-talker overlapping speech recognition remains challenging. 
Recent research revealed that ASR model’s encoder captures different levels of information with different layers -- the lower layers tend to have more acoustic information, and the upper layers more linguistic. 
This inspires us to develop a \textit{Sidecar} separator to empower a well-trained ASR model for multi-talker scenarios by separating the mixed speech embedding between two suitable layers.
We experimented with a wav2vec 2.0-based ASR model with a Sidecar mounted. By freezing the parameters of the original model and training only the Sidecar (8.7 M, 8.4\% of all parameters), the proposed approach outperforms the previous state-of-the-art by a large margin for the 2-speaker mixed LibriMix dataset, reaching a word error rate (WER) of 10.36\%; and obtains comparable results (7.56\%) for LibriSpeechMix dataset when limited training. 
\vspace{-0.05cm}

\end{abstract}
\vspace{-0.05cm}
\section{Introduction}
\label{sec:intro}

End-to-end automatic speech recognition (ASR) for common non-overlapping environments has made significant progress recently \cite{chan2016listen,gulati20_interspeech}. However, multi-talker (also known as multi-speaker) speech recognition, where overlapping may exist, still remains a challenge \cite{bai2021speaker}. 
There are two mainstream end-to-end paradigms that aim to tackle the challenge: (i) cascade architectures that jointly fine-tune speech separation and speech recognition modules \cite{settle2018end,li2021real}; and (ii) complete end-to-end models customized deliberately for multi-talker overlapping speech scenarios \cite{seki2018purely, zhang2020improving, chang2020end, tripathi2020end, kanda2020serialized, lu2021streaming, kanda2022streaming1}. However, the former approach may see performance degradation in modules' original domains, and the latter does not take full advantage of the readily available advancements made for single-talker ASR. This motivates us to find a low-cost and loose-coupling approach to adapt well-trained single-talker ASR models for multi-talker scenes without distorting the original model's parameters.

\begin{figure}[ht]
  \centering
  \centerline{\includegraphics[height=0.904\linewidth]{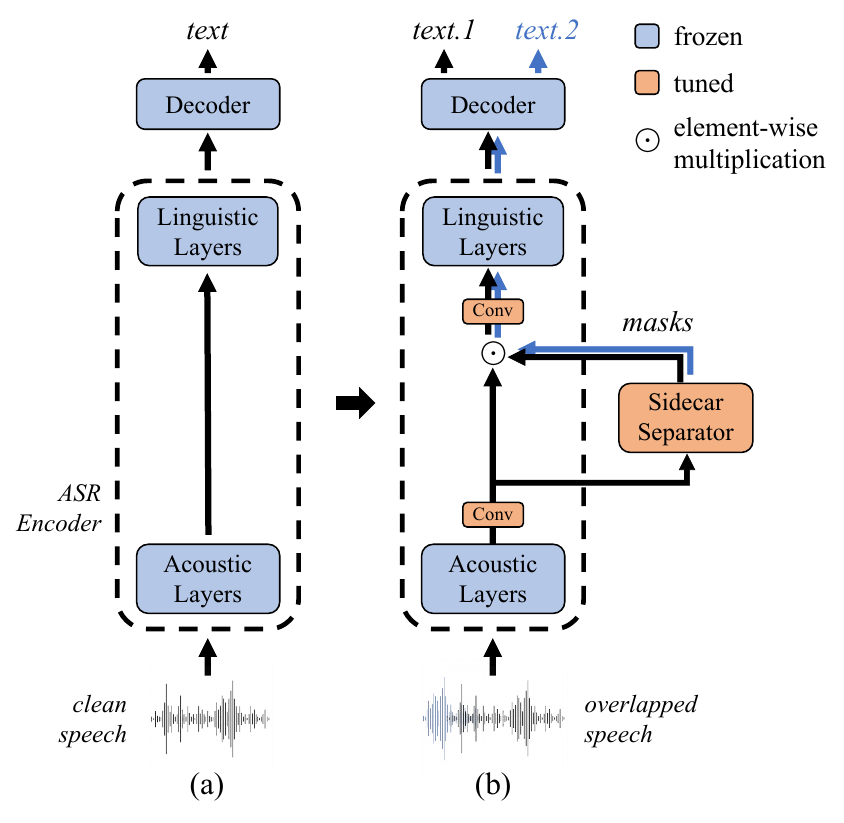}}
  \vspace{-0.5cm}
  \caption{(a) a well-trained single-talker ASR system; (b) the proposed strategy with a Sidecar separator mounted, taking 2 talkers as an example. On both sides of Sidecar, we add a convolutional layer to coordinate its input-overlapped and output-separated embeddings.}
  \label{fig:framework}
\end{figure}

Recent research investigated the information captured in the layers within the encoders of ASR models.
Shim et al. \cite{shim2021understanding} found that the transformer-based encoder extracts acoustic representations in its lower layers, and linguistic representations in the upper layers. Layer-wise analyses of self-supervised speech representation models, such as wav2vec 2.0 \cite{baevski2020wav2vec}, proved that they encode representations following an acoustic-linguistic hierarchy from lower to upper layers as well \cite{pasad2021layer, shah2021all}, which was further discussed in the context of neuroscience \cite{millet2022toward,vaidya2022self}. Similar findings have also been reported for CNN- / RNN-based models \cite{prasad2020accents, li2020does}.

Enlightened by the above findings, we assume that there exists a lower suitable location between the encoder's two layers where the multi-speaker overlapped acoustic embedding can be well-separated by drawing on speech separation techniques. In speech separation, research demonstrated that predicting masks for separation is usually superior to directly predicting separated embeddings \cite{wang2018supervised, wang2014training}. As a representative, the TasNet architecture \cite{luo2018tasnet} predicts masks in the time domain for mixed speech embeddings and achieves impressive results. Subsequently, Luo et al. \cite{luo2019conv} proposed the well-recognized Conv-TasNet, which predicts masks using a convolutional neural network, which made a further leap in performance.

Inspired by the recent analyses of ASR models and methodologies in speech separation, we introduce a promising strategy to adapt off-the-shelf well-trained ASR models to multi-talker scenarios. As shown in Fig. \ref{fig:framework}, we mount a \textit{Sidecar} separator between two suitable layers of a well-trained ASR model. On both sides of the Sidecar, there is a simple convolutional layer helping to coordinate the input-overlapping and output-separated embeddings of the Sidecar. The proposed approach has three key advantages:

\begin{itemize}
\item[$\bullet$] The approach is low-cost and loose-coupling for converting a well-trained single-talker ASR model to a multi-talker one, without complicated customization on the model structure or on the training scheme.
\item[$\bullet$] The original ASR model is well-trained and fixed, and only Sidecar (8.7 M, 8.4\% of all parameters) needs tuning, making the training feasible within limited time and GPU resources.
\item[$\bullet$] Experiments leveraging a wav2vec 2.0-based ASR model mounted with a Sidecar are conducted, achieving a WER of 10.36\% on 2-speaker LibriMix dataset and 7.56\% on LibriSpeechMix dataset with limited training.
\end{itemize}

Moreover, by visualizing Sidecar-predicted masks (Fig. \ref{fig:vi}),  we find that among channel dimensions different features encode different speakers' information. And in the time domain, there exist clear distinctions between the periods of speech for different speakers, and the periods of overlapping speech.

\section{Multi-talker ASR System with Sidecar}
\label{sec:methods}

The proposed methodology consists of three main components — a well-trained single-talker ASR model with parameters frozen, a Sidecar separator, and the training objective. 
As shown in Fig. \ref{fig:framework}, with the cooperation of two convolutional layers, Sidecar is plugged between two layers of ASR encoder and forms a multi-talker ASR system. No language models or lexicons are used in this work.

With permutation invariant training (PIT) \cite{yu2017permutation}, the model is optimized using connectionist temporal classification (CTC) loss \cite{graves2006connectionist}.
\vspace{-0.3cm}
\subsection{Well-trained single-talker ASR model}

A typical end-to-end ASR model contains an encoder to synthesize waveform or acoustic features into high-level representations, and a decoder to model the representations into language tokens. It often takes much time and effort to train an ASR model from scratch, let alone in multi-talker environments. With many off-the-shelf single-talker models already available, we try to reuse the single-talker model on multi-talker overlapping speech recognition.

Wav2vec 2.0 \cite{baevski2020wav2vec} is a well-recognized pre-trained speech representation model based on self-supervised learning (SSL), attracting interest in the field. Many ASR models taking wav2vec 2.0 as the encoder reported state-of-the-art performance \cite{chung2021w2v}.

In our experiments, we use a well-trained wav2vec 2.0 base-based ASR model, the same as used in \cite{baevski2020wav2vec}, which contains a CNN feature extractor, a Transformer encoder, and a fully-connected layer as the decoder.
Specifically, the model takes a waveform as input and extracts acoustic features with a 7-layer CNN feature extractor. After a linear projection, the features are fed into the encoder consisting of 12 layers of Transformer blocks for generating high-level representations. We obey the paradigm in \cite{baevski2020wav2vec} and only use a fully-connected layer as the decoder for letter-level prediction. 

We directly use the model parameters released by fairseq \cite{ott2019fairseq}, denoted as \emph{W2V-CTC} in the following parts.

\begin{figure}[ht]
  \centering
  \centerline{\includegraphics[width=0.90\linewidth]{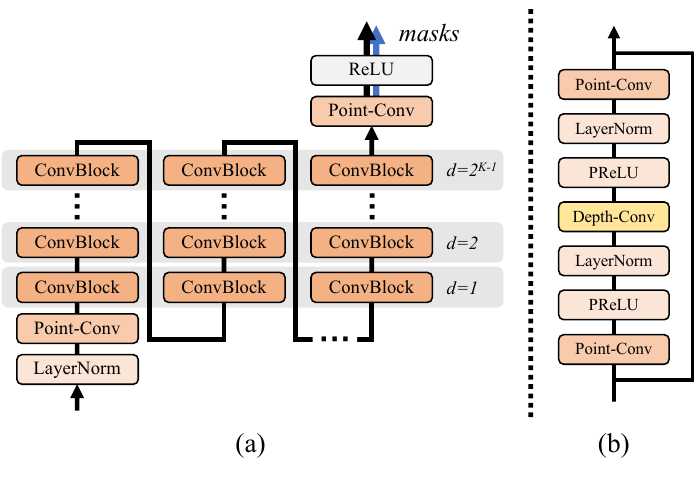}}
  \vspace{-0.3cm}
  \caption{(a) Sidecar structure; (b) Details in ConvBlock. Referring to Conv-TasNet \cite{luo2019conv}, the Sidecar consists of stacked 1-D dilated convolutional blocks. $d$ represents the dilation rate.}
  \label{fig:sidecar}
\end{figure}

\subsection{Sidecar separator}
Inspired by the findings that the ASR encoder captures more acoustic information in its lower layers and more linguistic information in the upper layers \cite{shim2021understanding, pasad2021layer,li2020does}, we propose using a Sidecar separator to address multi-talker speech recognition, drawing on methodologies in speech separation.

As shown in Fig. \ref{fig:sidecar}, similar to Conv-TasNet \cite{luo2019conv}, the Sidecar is a temporal convolutional network consisting of stacked 1-D dilated convolutional blocks, which allows the Sidecar to model the long-term dependencies of the acoustic embeddings while maintaining a small size. By ablation study, we plug the Sidecar into the most suitable position between two “acoustic” layers of the ASR encoder. Since the original ASR model is frozen, to alleviate the “transplant rejection”, we use a 3-kernel-size convolutional layer to filter Sidecar's input-mixed and output-separated embeddings, respectively.  

In the forward process, the preceding-layer-generated mixed speech embedding is filtered by a convolutional layer, and fed into the Sidecar to generate the speaker-dependent masks. Then, the filtered mixed speech embedding will be element-wise multiplied by the masks, and further adjusted by another convolutional layer to obtain the separated embeddings. The separated embeddings of different speakers will go in parallel through the rest of the model and be transcribed into text. This is technically implemented by concatenating the separated embeddings onto the batch dimension.

\subsection{Training objective}
We favor the use of permutation invariant training (PIT) with only ASR loss, which is CTC loss in this work. This can already achieve satisfactory performance. 

In addition to PIT-CTC loss, we also tried two reconstruction objectives: maximize scale-invariant signal-to-noise ratio (SI-SNR) or minimize mean squared error (MSE). Since the multi-talker dataset is simulated from single-speaker speeches, reconstruction loss aims to drive the predicted separated embeddings as close as possible to corresponding clean single-speaker embeddings. The clean single-speaker embeddings are generated by the transformer layer before where the Sidecar is plugged in, and the permutation of speakers is determined by PIT-CTC loss. 

Although introducing a reconstruction loss can provide a minor performance gain (Table \ref{tab:recon}), we do not recommend this. Because it requires input not only mixed speech but also clean single-speaker speeches, which significantly increases the computational burden.

\section{A Baseline System for control}
\label{baseline}
We attribute the effectiveness of this work to two aspects: the knowledge of the well-trained single-talker ASR model, and the Sidecar which can efficiently adapt the former to multi-talker scenarios by predicting speaker-dependent masks rather than their embeddings. The contribution of a well-trained model is intuitive, while the boost in performance provided by Sidecar can be indistinct.

Considering this, we designed a baseline system, which also leverages the same ASR model as our proposed approach, but directly predicts speaker-dependent speech embeddings like \cite{chang2020end}. 
Specifically, in the same position as Sidecar is in our proposed approach, the baseline model duplicates the preceding encoder layer to predict speaker-dependent embeddings. 
Except for the two duplicated layers, other parameters are frozen.

Note that, unlike our proposed approach, this Baseline does not maintain the property of keeping the original model parameters unchanged, because it fine-tunes the layers of the original model.

\section{Experimental Setup}
\label{sec:setup}

\subsection{Datasets}
\label{sec:date}

The experiments are performed on two benchmark datasets: \emph{LibriMix} \cite{cosentino2020librimix} and \emph{LibriSpeechMix} \cite{kanda2020serialized}.
Both of them are simulated from \emph{LibriSpeech} dataset but with different protocols.

\noindent
\textbf{LibriMix}. 
It is simulated with the mixtures of two or three speakers, in a clean or noisy environment. 
We focus on its 2-speaker-clean subset \emph{Libri2Mix-clean}. \emph{Libri2Mix-clean}'s training, development, and test set contain 270 hours, 11 hours, and 11 hours of 2-speaker's mixed speeches, respectively. The mixtures are made in a left-aligned style.
Therefore, the shorter source speech will be fully overlapped by the longer one from the beginning.

\noindent
\textbf{LibriSpeechMix}.
It only has standard dev and test sets.
We focus on the 2-speaker "\textit{dev-clean-2mix}" and "\textit{test-clean-2mix}" for validation and test.
The 2-speaker training set is homemade from the 960-hour LibriSpeech training dataset (LS-960) using the same protocol as \cite{kanda2020serialized}.
LibriSpeechMix randomly samples a delay time for a second utterance so the mixture is partially overlapping.

Compared with LibriSpeechMix, LibriMix has larger overlap ratios, which greatly challenges the model in separating overlaps.

\subsection{Model settings}
\noindent
\textbf{Well-trained single-talker ASR model}. 
In accordance with the paradigm in \cite{baevski2020wav2vec}, the used W2V-CTC model contains a CNN feature extractor, a Transformer encoder, and a fully-connected layer as the decoder.
We directly reuse the parameters released by Fairseq\footnote[1]{https://github.com/facebookresearch/fairseq/tree/main/examples/wav2vec} \cite{ott2019fairseq}, which was first pre-trained on unlabeled LS-960 with contrastive loss and diversity loss, then fine-tuned on labeled LS-960 with CTC loss. It reached a WER of 3.4\% on LibriSpeech test-clean dataset, and 8.9\% on test-other dataset.

\noindent
\textbf{Sidecar separator}. 
Referring to \cite{luo2019conv}, in our Sidecar separator, \textit{K} convolutional blocks with dilation rates 1, 2, 4, ..., 2$^{K-1}$ are repeated $R$ times. We take $K=8, R=3$, which performs better. We discard skip-connection paths of convolutional blocks, and change the final sigmoid activation to ReLU to fit our task. The Sidecar uses 128 bottleneck-channels, and 768 input- / output- channels.  Ablation experiments (Section \ref{sec:ablation}) have been conducted to explore the most suitable Sidecar location. As a result, we plug the Sidecar right after the second transformer layer and before the third, which gave the best performance. 
With W2V-CTC frozen, it only has 8.7 M (8.4\% of all parameters, about half of the Baseline) for tuning.

\noindent
\textbf{Training settings}. 
We optimize the proposed model and Baseline using a 2e-4 learning rate with a three-stage schedule and Adam optimizer, for at most 100 k updates. It takes about 7 hours for models' convergence with 8 NVIDIA V100 GPUs, thanks to Sidecar's small size and the ejection start provided by the well-trained model.

In the following, we denote the proposed model as \textit{W2V-Sidecar}.

\section{Results and Discussion}
\label{sec:results}

\subsection{Results On LibriMix dataset}

\begin{table}[!htbp]
  \caption{Comparison of different systems on \emph{LibriMix}. Evaluated by WER (\%). “Transf.” refers to “Transformer” and “ft.” refers to “fine-tune the whole model”.}
  \vspace{-0.2cm}
  \label{tab:base-libmix}
  \centering
  \setlength{\tabcolsep}{4mm}{
  \begin{tabular}{lcc}
\\\toprule
\textbf{System} & \textbf{Dev}  & \textbf{Test} \\
\hline
\rule{-2.2pt}{9pt}

(a) PIT-Transf. \cite{li2021real}   &   26.58  &   26.55    \\
(b) Conditional Conformer \cite{guo2021multi}   &   24.50  &   24.90     \\
(c) Conv-TasNet + Transf. \cite{li2021real}   &   21.00  &   21.90   \\
(d) DPRNN-TasNet + Transf. \cite{li2021real}   &   15.30  &   14.50   \\
\hline
\rule{-2.2pt}{9pt}
(e) Baseline (proposed)&   11.60   &   12.27    \\
\hline
\rule{-2.2pt}{9pt}
(f) W2V-Sidecar (proposed)  &  \textbf{9.76}  &   \textbf{10.36}    \\
(g) W2V-Sidecar-ft. (proposed)&   \textbf{7.68}  &   \textbf{8.12}   \\
\bottomrule
\end{tabular}}
\vspace{-0.2cm}
\end{table}

\begin{table}[!htbp]
  \caption{Comparisons of different systems on \emph{LibriSpeechMix}. Evaluated by WER (\%). “-” refers to “not reported” and “ft.” refers to “fine-tune the whole model”.}
  \vspace{-0.2cm}
  \label{tab:base-spemix}
  \centering
  \begin{threeparttable}
  \setlength{\tabcolsep}{4.5mm}{
  \begin{tabular}{lcc}
\\ \toprule
\textbf{System} & \textbf{Dev} &\textbf{Test}\\
\hline
\rule{-2.2pt}{9pt}
(a) PIT-BiLSTM \cite{kanda2020serialized}   &   -   &   11.10   \\
(b) SOT-BiLSTM \cite{kanda2020serialized}   &   -    &  11.20   \\
(c) SURT-non-streaming \cite{lu2021streaming}   &   -    &   7.20\textsuperscript{\dag}   \\
(d) SOT-transf. \cite{kanda2021end}   &   -   &   5.30\textsuperscript{\dag}   \\
\hline
\rule{-2.2pt}{9pt}
(e) Baseline  (proposed)&   9.50    &   9.41   \\
\hline
\rule{-2.2pt}{9pt}
(f) W2V-Sidecar  (proposed)&   7.76   &   7.56    \\
(g) W2V-Sidecar-ft. (proposed)&   6.01   &   5.69  \\
\bottomrule
\end{tabular}}
\begin{tablenotes}
\item \textsuperscript{\dag}With heavier training data.
\end{tablenotes}  
\end{threeparttable}
\end{table}

We compared different systems on the two benchmark datasets.
The corresponding results are shown in Table \ref{tab:base-libmix} and Table \ref{tab:base-spemix}.

For 2-speaker LibriMix (Table \ref{tab:base-libmix}), the designed Baseline (e) for control already outperforms previous methods by a large margin.
We attribute this improvement to the knowledge of the well-trained model.
Then, by comparing the proposed W2V-Sidecar (f) with Baseline (e), we find the introduction of Sidecar further boosts the WER with even less trainable parameters (8.7 M, 8.4\% of all parameters, about half of Baseline). This confirms that predicting masks is more effective than directly separating embeddings as discussed in \cite{wang2014training}.
Besides, the Sidecar serves in a plug-and-play style without distorting the original parameters. 
This low-coupling property allows the model to be more flexible for multiple scenarios.

Moreover, as an option, we also train a W2V-Sidecar-ft (g), which fine-tunes all model parameters. The training settings are the same as (e). Since the model is initialized with well-trained parameters, the W2V-Sidecar-ft's convergence is also fast. Not surprisingly, the model achieves even more impressive results.

\vspace{-0.2cm}
\subsection{Results On LibriSpeechMix dataset}

For the comparison on the LibriSpeechMix dataset, we only focus on those non-streaming models with no additional auxiliary such as  speaker labels or additional training datasets.

As shown in Table \ref{tab:base-spemix}, our Baseline (e) achieves better performance than (a) and (b) even with fewer training efforts. As a similar trend, the Sidecar (f) brings a further performance boost. And as an option, the W2V-Sidecar-ft (g) reaches a better result at the cost of losing the loose-coupling property. 
We also list the results of Systems (c) and (d), which with different setups, for a comprehensive comparison. They gain further improvements from their significantly heavier training efforts.

Compared with (a)-(d), the proposed systems (e)-(f) are trained efficiently with only 7 hours using 8 GPUs. Moreover, systems (a), (b), and (d) have larger model sizes than the proposed W2V-Sidecar (94.4 M frozen + 8.7 M trainable).

\subsection{Visualization of Sidecar predicted masks}

To better understand what Sidecar has learned, we investigate its generated masks with visualization. For a better view, we perform element-wise softmax on the two masks derived from a mixed speech embedding to highlight pairwise differences. This process produces two essentially identical matrices, and we just take one. Then we normalize each channel (or feature) using its mean and standard deviation along the time steps to avoid swamping those channels which with minor differences between the two masks. Afterward, we reorder the channel dimension according to hierarchical clustering based on the pairwise distances of the channels.

We randomly take three typical cases as examples: an almost non-overlapped case, a partially overlapped case, and a case in which the shorter speech is fully overlapped. We find interesting clues from their masks' visualization (Fig. \ref{fig:vi}). The masks show strong temporal correlations with the input speech, and distinctly tell the mixture boundaries.
This indicates that different feature sub-spaces (or channel groups) capture different speakers' information, so the speaker boundaries emerge when close channels are clustered.

\vspace{-0.14cm}

\subsection{Ablation studies}
\label{sec:ablation}

\noindent
\textbf{Sidecar location}.
We explored the optimal location for the Sidecar. 
We mark the locations by the transformer blocks index. E.g., location \textbf{0} is right before the first block; location \textbf{1} is between the first and second blocks, etc. 
The results are summarized in Table \ref{tab:ablation}. 
Among the locations, the result of location \textbf{2} exceeds the previous layers, and deeper locations show a dramatic decrease in results.

This trend is mutually supportive of existing layer-wise analysis research \cite{shim2021understanding, pasad2021layer}, and aligns with our previous hypothesis: the separation is better performed on acoustic-related representations, which contain sufficient low-semantics information to catch phonetic-level differences. We argue that location \textbf{0} is too close to the raw input, where meaningful phonetic-level representations have yet to be well-synthesized. Meanwhile, speaker information matters, as the discussion about Fig. \ref{fig:vi}. However, if the ASR encoder goes deeper, the speaker information will be eliminated. As a result, location \textbf{2} is a compromised scale in semantics.

\noindent
\textbf{Reconstruction loss}.
As reconstruction loss plays a dominant role in speech separation tasks, a natural question is whether introducing reconstruction loss can help our task.
In this part, we explore introducing MSE or SI-SNR loss to the Sidecar strategy.
In Table \ref{tab:recon}, our experiments show that introducing reconstruction loss can make slight improvements. However, the computational burden is significantly increased because it requires not only mixed speech input but also clean single-talker recordings.
We argue that adding a constraint on low-level embeddings for such a high-semantic task may not be very helpful because the mapping can be an ill-posed problem.

\vspace{-0.14cm}

\subsection{Limitations and future work}
\begin{figure}[ht]

  \centering
  \centerline{\includegraphics[width=0.994\linewidth]{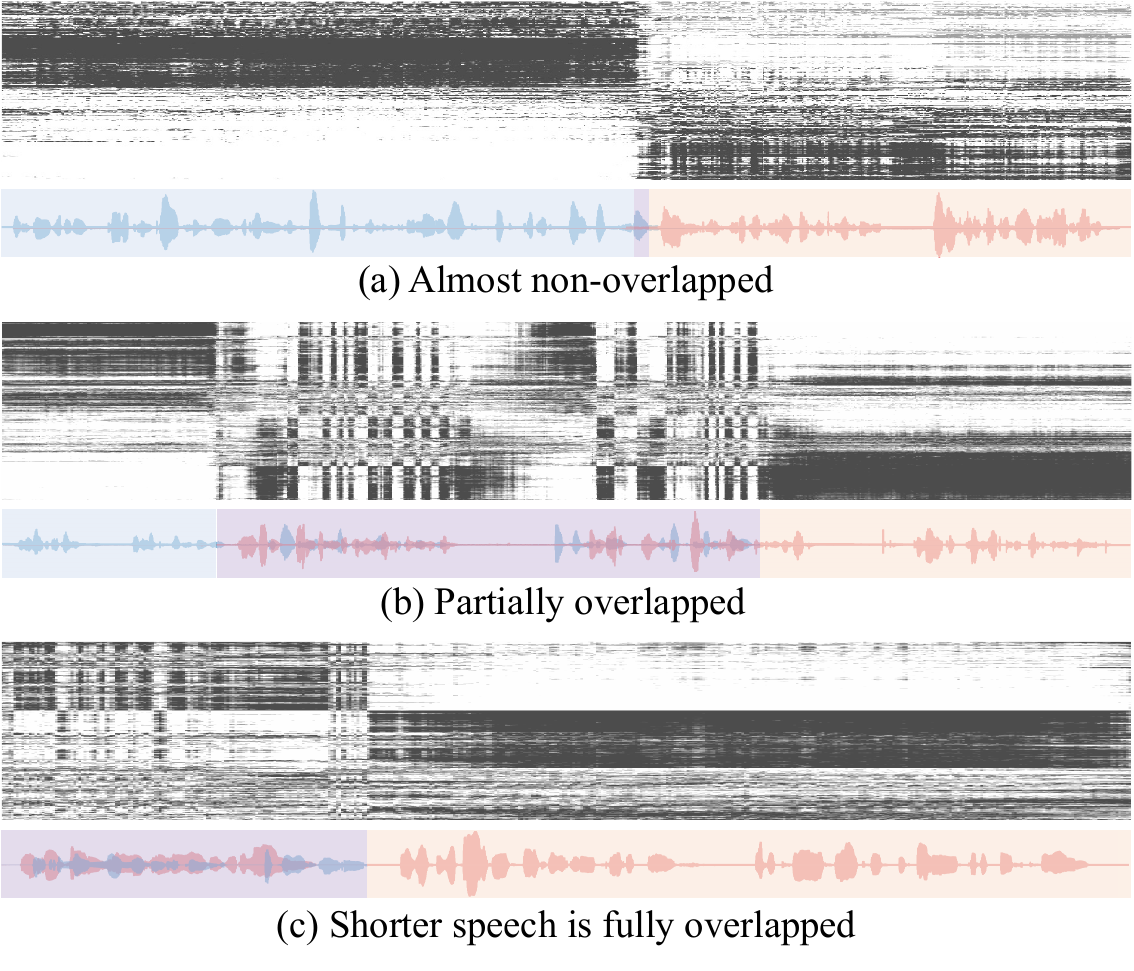}}
  \vspace{-0.5cm}
  \caption{Visualization of generated masks and input waveforms. We use different colors to distinguish speakers. Purple represents overlapping. The horizontal of masks is time dimension, and the vertical is channel dimension. Sidecar encodes speaker information with different channels and indicates clear distinctions in time domain.}
  \label{fig:vi}
  \vspace{-0.1cm}
\end{figure}

This work has several limitations. 
First, although we used PIT as our training scheme, our strategy also naturally fits serialized output training (SOT), which usually needs to train from scratch. We are interested in whether Sidecar can accelerate SOT's training with a well-trained ASR model.
Second, according to Fig. \ref{fig:vi}, Sidecar explicitly encodes speaker information. We are excited about the prospects of its application to speech diarization, especially combined with SOT.
Third, we only implement Sidecar on the ASR task. We also expect its applicability to other downstream tasks when overlapping exists.
Finally, to adopt a generally popular speech representation model, we only use wav2vec 2.0-based model. We will try the Sidecar on other SSL or non-SSL models in the future.

\begin{table}[!htbp]
  \caption{Ablation study on Sidecar location, evaluated by WER (\%).}
    \vspace{-0.3cm}
  \label{tab:ablation}
  \centering
    \setlength{\tabcolsep}{1mm}{
  \begin{tabular}{cccccccccc}
\\\toprule[1pt]
&\multicolumn{8}{c}{Location}\\
\cmidrule(r){2-9}
  LibriMix & 0 & 1& 2 & 3& 4& 6& 9 & 12 \\
  \hline
\rule{-2.2pt}{9pt}
  Dev & 12.18 & 11.22 & \textbf{9.76} & 12.06 & 16.14 & 30.03 & 56.38 & 61.78 \\
  Test & 13.01 & 11.87 & \textbf{10.36} & 12.65 & 16.88 & 30.32 & 57.11 & 62.72 \\
\bottomrule[1pt]
\end{tabular}}
\vspace{-0.25cm}
\end{table}

\begin{table}[!htbp]
  \caption{Ablation study on using reconstruction loss, by WER (\%).}
    \vspace{0.1cm}
  \label{tab:recon}
  \centering 
  \setlength{\tabcolsep}{4mm}{
  \begin{tabular}{lcccc}

    \toprule
    & \multicolumn{2}{c}{LibriMix} & \multicolumn{2}{c}{LibriSpeechMix} \\
    \cmidrule(r){2-3} \cmidrule(r){4-5}
     & Dev & Test   &    Dev & Test     \\
    \hline
    \rule{-2.2pt}{9pt}
    W2V-Sidecar            &   9.76    & 10.36   & 7.76  &   7.56 \\
      \quad\quad w/ SISNR   & \textbf{9.69} & \textbf{10.16} & \textbf{7.43} & \textbf{7.20}  \\
      \quad\quad w/ MSE  &   9.74    &  10.32  & 7.90    &   7.34  \\
    \bottomrule
    \end{tabular}
    }
    \vspace{0.2cm}
\end{table}

\section{Conclusion}
\label{sec:conclusion}

\vspace{-0.13cm}
Inspired by the findings that ASR encoder captures more acoustic
representations in its lower layers and more linguistic in the upper layers, we propose plugging a Sidecar separator into a well-trained single-talker ASR model and converting it to a multi-talker one. The original ASR model is frozen, and only 8.4\% of all parameters need tuning. With efficient training, the proposed method outperforms previous state-of-the-art by a large margin on the 2-speaker mixed LibriMix dataset, reaching a WER of 10.36\% dataset; and comparable results (7.56\%) on the LibriSpeechMix dataset.

Visualizations of Sidecar-predicted masks indicate that in channel dimension, different features encode different speakers’ information. And in the time domain, there exist significant distinctions between different speaker speech periods and overlapping periods.

\vspace{-0.14cm}

\section{Acknowledgements}
\label{sec:ack}
\vspace{-0.13cm}
\thanks{This research is partially supported by the HKSARG Research Grants Council’s Theme-based Research Grant Scheme (Project No. T45-407/19N) and by the CUHK Stanley Ho Big Data Decision Analytics Research Centre. }

\vfill\pagebreak

\bibliographystyle{IEEEbib}
\bibliography{refs}

\end{document}